\documentclass[%
reprint,
nofootinbib,
 amsmath,amssymb,
aps,
prl,
]{revtex4-1}

\usepackage{tikz}
\usetikzlibrary{matrix,arrows,decorations.pathmorphing,decorations.text,decorations.pathreplacing}
\usepackage{amsmath,amssymb,euscript,array,mathrsfs,appendix,ctable}

\def\mf{{\mathfrak f}}
\def\mfp{{\mathfrak p}}

\newcommand{\Tr}{\operatorname{Tr}}

\newcommand{\AD}{\operatorname{ad}}

\def\B0{{\boldsymbol 0}}

\def\mg{\mathfrak{g}}

\def\Tr{{\rm Tr}}

\def\SU{\text{SU}}
\def\SO{\text{SO}}

\def\U{\text{U}}

\def\Dbarslash{\,\,{\raise.15ex\hbox{/}\mkern-12mu {\bar D}}}
\def\Dslash{\,\,{\raise.15ex\hbox{/}\mkern-12mu D}}
\def\delslash{\,\,{\raise.15ex\hbox{/}\mkern-9mu \partial}}
\def\delbarslash{\,\,{\raise.15ex\hbox{/}\mkern-9mu {\bar\partial}}}

\newcommand{\MAT}[1]{\begin{pmatrix} #1\end{pmatrix}}

\newcommand{\EQ}[1]{\begin{equation}\begin{split} #1
\end{split}\end{equation}}

\newcounter{Part}
\newcommand{\Part}{\refstepcounter{Part}\vspace{0.5cm}{\bf \arabic{Part}. }}

\begin{document}

\preprint{APS/123-QED}

\title{Symplectic Deformations of Integrable Field Theories and AdS/CFT}

\author{Timothy J. Hollowood}
\email{t.hollowood@swansea.ac.uk}
\affiliation{Department of Physics, Swansea University, Swansea, SA2 8PP, U.K.}

\author{J. Luis Miramontes}
\email{jluis.miramontes@usc.es}
\affiliation{Departamento de F\'\i sica de Part\'\i culas and IGFAE,
Universidad de Santiago de Compostela, 15782 Santiago de Compostela, Spain}

\date{\today}

\begin{abstract}Relativistic integrable field theories like the sine-Gordon equation have an infinite set of conserved charges.
In a light-front formalism these conserved charges are closely related to the integrable modified KdV hierarchy at the classical level. The latter hierarchy admits a family of symplectic structures which we argue can be viewed as deformations of the relativistic sine-Gordon symplectic structure. These deformed theories are integrable but no longer relativistic and the basic excitations of the theory, the solitons, have an interesting non-relativistic dispersion relation that in a certain limit becomes the dispersion relation of dyonic giant magnons of string theory in the AdS/CFT correspondence. 
We argue that the deformed classical theories can be lifted to quantum theories when the sine-Gordon theory is embedded in a larger theory that describes the string world-sheet sigma model in $\text{AdS}_5\times S^5$.
\end{abstract}

\pacs{Valid PACS appear here}
\maketitle

\pgfdeclarelayer{background layer} 
\pgfdeclarelayer{foreground layer} 
\pgfsetlayers{background layer,main,foreground layer}

\Part The sine-Gordon theory is the most iconic relativistic integrable field theory in $1+1$ dimensions. It even plays a role as a limited sector of the integrable structure that lies behind the hidden Integrability of the AdS/CFT correspondence. In this case, classical integrability can be seen explicitly on the world-sheet of the string, and the sine-Gordon theory describes the sector where the string moves in $\mathbb R\times S^2\subset\text{AdS}_5\times S^5$. 

What is interesting is that it provides a very simple arena to describe certain integrable deformations of the string world-sheet sigma model that potentially yield deformations of the complete AdS/CFT duality.
The approach in this letter is complementary to the approach of \cite{Delduc:2012qb,Delduc:2012mk,Delduc:2012vq,Delduc:2013fga,Delduc:2013qra} who consider integrable deformations of the string world-sheet sigma model directly in the Hamitonian formalism. Here, we shall follow \cite{Mikhailov:2006uc,Mikhailov:2007xr,Schmidtt:2011nr} and work in a light-front formalism that makes the relation with the well-known integrable hierarchies and the soliton solutions more concrete. 

\Part
The sine-Gordon equation takes the form
\EQ{
\partial_+\partial_-\phi+\sin\phi=0\ ,
}
where $x^\pm= t\pm x$ are light-cone coordinates.
It is famously integrable since there exists an infinite series of conserved charges $Q^{(s)}$ of odd spin $s$.  The pair \mbox{$p_\pm=Q^{(\pm1)}$} are the components of the energy-momentum vector. All these charges Poisson commute in the classical theory:
\EQ{
\{Q^{(s)},Q^{(s')}\}=0\ .
}
It is useful, in the following, to work in a light-front formalism on surfaces $x^-=\text{const}$. The Poisson bracket is then 
\EQ{
\{\phi(x^+,x^-),\partial_+\phi(y^+,x^-)\}=\delta(x^+-y^+)\ .
\label{u78}
}

The conserved charges generate Hamiltonian symmetries that are conveniently written in terms of  $q=\partial_+\phi$
\EQ{
\frac{\partial q}{\partial t^{(s)}}=\{q,Q^{(s)}\}\ ,
}
where $x^\pm\equiv t^{(\pm1)}$. The flow $t^{(3)}$ is identified with the mKdV equation
\EQ{
\frac{\partial q}{\partial t^{(3)}}=-\partial_+^3q-\frac32q^2\partial_+q\ ,
}
while the other positive flows $t^{(s)}$, $s>0$, are polynomial in $q$ and its $\partial_+$-derivatives and give the whole mKdV hierarchy of integrable equations. In contrast, the negative flows $t^{(s)}$, $s<0$, turn out to be non-local. The first non-trivial one is
\EQ{
\frac{\partial q}{\partial t^{(-3)}}=&\cos\phi\,\partial_+^{-1}\left(\cos\phi\,\partial_+^{-1}\sin\phi\right)\\[5pt]
&
+\sin\phi\,\partial_+^{-1}\left(\sin\phi\,\partial_+^{-1}\sin\phi\right)\ .
}
Written in terms of $\tilde{q}=\partial_-\phi=-\partial_+^{-1}\sin\phi$, the negative flows give another copy of the mKdV hierarchy.

It is a key property of such an integrable hierarchy that it can be described in terms of a multi-Hamiltonian structure~\cite{Magri:1977gn,Fokas:1997,Mikhailov:2005sy}, so that
the same flows can be written in terms of an infinite set of other Poisson brackets
\EQ{
\frac{\partial q}{\partial t^{(\pm s)}}&=\{q,Q^{(\pm s\pm 2n)}\}_{\mp n}= \theta_{\mp n} \frac{\delta Q^{(\pm s\pm 2n)}}{\delta q}\ ,
\label{tff}
}
with both $s$ and $s+2n>0$. 
Here, $\theta_n$ are the non-local differential operators
\EQ{
\theta_n= (-1)^n(\partial_+^2+\partial_+ q\partial_+^{-1}q)^{n}\partial_+ 
= \theta_0 \theta_{-n}^{-1} \theta_0\,.
}
Notice that in~\eqref{tff} there are two separate towers, since
\EQ{
\{q,Q^{(\pm s)}\}_{\mp n}=0\quad \text{for}\quad s-2n<0
}
for each value of $s>0$.
All the Poisson brackets are {\it coordinated\/}, meaning that any linear combination is also a valid Poisson bracket and so satisfies the Jacobi identity. For $n\not=0$, they are non-local and their rigorous description is still an open problem~(see~\cite{Maltsev:2001,DeSole:2013} and the references therein). 
The original symplectic structure \eqref{u78} is $\{\ ,\ \}_{0}$, which is the only relativistic invariant one, and $\theta_1$ provides the, so-called, second Hamiltonian structure~\cite{Magri:1977gn,Fokas:1997}. 

The combination 
\EQ{
\theta= -\theta_1+ 2\theta_0 -\theta_{-1}
\label{spb}
}
gives the Poisson bracket of the gauge-fixed world-sheet sigma model of the bosonic string moving on ${\mathbb R}\times S^2$ \cite{Mikhailov:2005sy,Mikhailov:2006uc,Mikhailov:2007xr}, which is non-relativistic once the Virasoro constraints are imposed. One can now imagine deforming the theory by changing the symplectic structure. In particular we shall be interested in the deformation inspired by~\eqref{spb}
\EQ{
\{\Phi,\Psi\}^{(0)}&\longrightarrow \{\Phi,\Psi\}_\sigma=\kappa\left(-\sigma^{-2}\{\Phi,\Psi\}_{1}\right.
\\ &
\hspace{-0.5cm}
\left.+\big(1+\sigma^{-4}\big)\{\Phi,\Psi\}_{0}-\sigma^{-2}
\{\Phi,\Psi\}_{-1}\right)\ ,
\label{def}}
where $\sigma\in[1,\infty]$ and $\kappa$ is an overall normalization. It is clear that for finite $\sigma$ the deformed theory will not be relativistic either.
In this new theory one can ask what are the energy and momentum. 
We can identify these as the generators of space-time translations 
\EQ{
\partial_\pm q=\{q,p_\pm^{\sigma}\}_\sigma\ ,
}
giving
\EQ{
p^\sigma_\pm=\kappa^{-1}\sum_{n=0}^\infty \sigma^{-2n} Q^{(\pm 2n\pm1)}\ .
}

\Part The picture above generalises to a class of generalised sine-Gordon (GSG) theories that are associated to any symmetric space $F/G$ \cite{Miramontes:2008wt}. They describe the Pohlmeyer reduction of sigma models with $F/G$ as target. A symmetric space is naturally associated to an involution $\sigma_-$ of the Lie algebra $\mathfrak f$ that provides the decomposition into eigenspaces $\mf=\mg\oplus\mfp$, with $\sigma_-(\mg)=\mg$ and $\sigma_-(\mfp)=-\mfp$. We can then construct a twisted affine loop algebra by associating each elements of $\mg$  and $\mfp$ with appropriate powers of an arbitrary parameter~$z$:
\EQ{
\hat\mf=\bigoplus_n\Big(\mg z^{2n}\oplus\mfp z^{2n+1}\Big)\ .
}

The basic field is $\gamma\in G\subset F$. The equation-of-motion can be written in Lax form as an $\hat\mf$-valued connection with light-cone components
\EQ{
{\cal L}_+(z)&=\partial_++\gamma^{-1}\partial_+\gamma-z\Lambda\ ,\\ {\cal L}_-(z)&=
\partial_--z^{-1}\gamma^{-1}\Lambda\gamma\ ,
}
where $z$ is the spectral parameter and $\Lambda$ is a constant element of $\mfp$. The equation-of-motion is then the flatness condition
\EQ{
[{\cal L}_+(z),{\cal L}_-(z)]=0\ .
\label{n22}
}

The theory can be formulated in a manifestly relativistic way as a gauged WZW model for the field $\gamma\in G\subset F$ gauged with respect to a subgroup $H\subset G$ defined as the centraliser of $\Lambda$ acting as $\gamma\to h\gamma h^{-1}$. The WZW model is then perturbed by the potential term
$\Tr(\gamma^{-1}\Lambda\gamma\Lambda)$. The level $k$ of the WZW term is the discrete coupling of the theory. In the on-shell gauge $A_\mu=0$ the equation-of-motion is precisely the flatness condition \eqref{n22}~\cite{Bakas:1995bm}.

The sine-Gordon theory itself is the example $\SO(3)/\SO(2)$ with
\EQ{
\gamma={\small\MAT{1&0&0\\ 0&\cos\phi&\sin\phi\\ 0&-\sin\phi&\cos\phi}}\ ,\quad\Lambda={\small\MAT{0&-1&0\\ 1&0&0\\ 0&0&0}}\ .
\label{j23}
}
In this case there is no WZ term and the coupling $k$ needs not be quantised.
The theory has a set of conserved quantities $Q[b]$ for each element of the affine algebra such that 
$b\in\text{Cent\;}(\text{Ker}\,\AD_\Lambda)$. For sine-Gordon theory
one has $Q^{(2n+1)}=Q[z^{2n+1}\Lambda]$. 

The next simplest theory is associated to the symmetric space $S^3=\SO(4)/\SO(3)$ and is constructed via the obvious generalisation of \eqref{j23}. This is the complex sine-Gordon theory. In this case, embedding $\SO(3)$ in the bottom right-hand corner of the 4-dimensional defining representation of $\SO(4)$, $\text{Ker}\AD_\Lambda$ contains 2 elements
\EQ{
\Lambda={\small\frac12\MAT{0&-1&0&0\\ 1&0&0&0\\ 0&0&0&0\\ 0&0&0&0}}\ ,\quad\tau={\small\MAT{0&0&0&0\\ 0&0&0&0\\ 0&0&0&-1\\ 0&0&1&0}}\ .
\label{j24}
}
Here, $\tau$ is the generator of $H=\SO(2)$.
This means that there are now two infinite series of commuting conserved quantities $Q[z^{2n+1}\Lambda]$ and $Q[z^{2n}\tau]$. The pair $Q[z^{\pm1}\Lambda]$ are once again identified with the light-cone components $p_\pm$, up to a scaling. But now there is a new spinless charge $Q[\tau]$ which is simply the $\SO(2)$ charge of the complex sine-Gordon theory.

The whole story of the Poisson brackets goes through exactly as for the sine-Gordon theory~\cite{Schmidtt:2011nr}. When formulated on the field $q=\gamma^{-1}\partial_+\gamma$, one can write
\EQ{
\{\Phi,\Psi\}_{n}=-\int dx^+\,\Tr\,\Big(\frac{\delta \Phi}{\delta q}\big(\AD_\Lambda D_+^{-1}\big)^{-2n}
D_+\frac{\delta \Psi}{\delta q}\Big)\ ,
}
where $D_+\lambda=[\partial_++q,\lambda]$. For $SO(3)/SO(2)$ these Poisson brackets reduce to those given in~\eqref{tff} up to an overall factor of $\frac12$.
Just as in the sine-Gordon case, one can define a family of symplectic structures $\{F,G\}_\sigma$, and we normalize it with $\kappa=4\pi[k(1-1/\sigma^2)]^{-1}$.
In the limit \mbox{$\sigma\to\infty$} with $k$ fixed we recover the Poisson bracket of the GSG theory. However, in the alternative limit
$\sigma \to1$ as $k\to\infty$ with $g=k(\sigma-\sigma^{-1})/4\pi$ fixed,
one finds the Poisson bracket of the gauge-fixed bosonic string sigma model on ${\mathbb R}\times
F/G$ where $g$ is the sigma model coupling. 
Note that in this case the non-local looking form of the Poisson bracket is an artefact of the gauge fixing procedure \cite{Mikhailov:2006uc}.

In the deformed theory the energy, momentum and $\U(1)$ charge become \EQ{
E&=\frac k{4\pi}(\sigma+1/\sigma)
\sum_{n\in\mathbb Z}\sigma^{-|2n+1|}Q[z^{2n+1}\Lambda]\ ,\\
p&=\frac k{4\pi}(\sigma-1/\sigma)\sum_{n\in\mathbb Z}\text{sign}(n)\sigma^{-|2n+1|}Q[z^{2n+1}\Lambda]\ ,\\
\EuScript Q&=\frac{k}{4\pi}\sum_{n\in\mathbb Z}\sigma^{-2|n|}Q[z^{2n}\tau]\ .
\label{p90}
}
so that $p^\sigma_\pm =(\xi E \pm p)/2$, with $\xi=(\sigma^2-1)/(\sigma^2+1)$.

\Part This identification is supported by taking a soliton of the GSG theory and evaluating its energy, momentum and charge \cite{Hollowood:2009tw,Hollowood:2010dt,Hollowood:2013oca}. A soliton depends on the complex parameters $z^\pm=e^{-\theta\pm i\alpha}$:
\EQ{
&Q[z^{\pm(2n+1)}\Lambda]=\frac{4\sin((2n+1)\alpha)}{2n+1}e^{\mp(2n+1)\theta}\ ,\\
&Q[z^{\pm2n}\tau]=\frac{4\sin(2n\alpha)}{n}e^{\mp2n\theta}\ ,\qquad n\geq0\,,
}
and so
\EQ{
\frac E{\sigma+1/\sigma}\pm\frac p{\sigma-1/\sigma}
=\frac k{2\pi i}\log\Big[\frac{z^\mp-\sigma^{\pm1}}{z^\mp+\sigma^{\pm1}}\cdot
\frac{z^\pm+\sigma^{\pm1}}{z^\pm-\sigma^{\pm1}}\Big]
\label{ft5}
}
 and 
\EQ{
\EuScript Q=\frac{k}{2\pi i}\log\Big[\frac{(\sigma z^+)^2-1}{\sigma^2- (z^+)^2}\cdot\frac{\sigma^2- (z^-)^2}{(\sigma z^-)^2-1}\Big]\ .
}
When the soliton is semi-classically quantized using the Bohr-Sommerfeld method, the charge $\EuScript Q$ is an integer and so this fixes $\alpha=\alpha(\theta)$.

One can verify that $\tanh\theta=\partial E/\partial p$ which identifies $\theta$ as the rapidity. The dispersion relation of the solitons then follows as \cite{Hoare:2013ysa}
\EQ{
\sin^2\Big(\frac{\xi E}{4g}\Big)-\xi^2\sin^2\Big(\frac p{4g}\Big)=(1-\xi^2)\sin^2\Big(\frac{\pi\EuScript Q}{2k}\Big)\ .
\label{disp}
}
Note that, written in this way, it can be presented as an exact equation by giving the exact $\sigma(g,k)$ below.

In the relativistic limit $\sigma\to\infty$ this gives the usual relativistic dispersion relation of the GSG theory
\EQ{
E^2-p^2=\frac{4k^2}{\pi^2}\sin^2\Big(\frac{\pi\EuScript Q}{2k}\Big)\ .
}
In the string sigma model limit, $\sigma\to1$ as $k\to\infty$, we have
\EQ{
E^2={\EuScript Q}^2+16g^2\sin^2\Big(\frac p{4g}\Big)\ ,
}
which is the dispersion relation of the dyonic giant magnons \cite{Dorey:2006dq}.

\Part In order to complete the relation to the  AdS/CFT one needs to add fermions. This is done by taking $F/G$ to be a semi-symmetric space. In the case of $\text{AdS}_5\times S^5$ the semi-symmetric space is \cite{Grigoriev:2007bu,Hollowood:2011fq}
\EQ{
\text{PSU}(2,2|4)/\text{Sp}(2,2)\times\text{Sp}(4)
\label{kqq}
}
and the GSG theory is then an ${\cal N}=(8,8)$ supersymmetric theory with $H=\SU(2)^4$ R-symmetry~\cite{Hollowood:2011fq,Goykhman:2011mq}.

In this context there is an exact conjecture for the S-matrix of the deformed theory based on a quantum group
deformation of the magnon S-matrix of the string sigma model with deformation parameter $q=\exp\big(i\pi/k\big)$ \cite{Hoare:2011wr,Hoare:2012fc}.
The dispersion relation of the magnon/soliton excitations are precisely given in the quantum theory by \eqref{disp} with integer charges $\EuScript Q$ but with the exact relation 
\EQ{
\sigma-\sigma^{-1}=4g\sin(\pi/k)\ .
}
These excitations transform in particular representations of the quantum supergroup $U_q(\mathfrak{psu}(2|2))^{\times2}$ which includes $U_q(H)$ as its bosonic subgroup.

It is possible to check the S-matrix ansatz in the semi-classical limit, that is $g,k\to\infty$ with fixed ratio $g/k$. In addition, the states with large charge, which are those where $\EuScript Q/k$ is fixed as $k\to\infty$, are realised as semi-classical soliton states in the field theory. The S-matrix of these states can then be compared
against the classical scattering of the solitons of the deformed GSG theory using the Jackiw-Woo formula \cite{Jackiw:1975im}
\EQ{
S(E)\thicksim\exp\left[i\int^EdE'\,\Delta t(E')\right]\ ,
}
where $\Delta t(E)$ is the classical time delay experienced by one soliton as it moves through another. In the deformed theory, the time delay is equal to that in the GSG theory because the equation-of-motion is independent of the deformation, but the energy must be the deformed quantity~\eqref{p90}. The soliton time delays can be extracted from the exact two soliton solutions constructed by the dressing method in \cite{Hollowood:2010dt} as will be shown elsewhere.

\Part To compare with the work of \cite{Delduc:2012qb,Delduc:2012mk,Delduc:2012vq,Delduc:2013fga,Delduc:2013qra}, note that
the deformation of the string sigma model considered in those references 
corresponds to taking $k$ imaginary. This means that $q=e^{-\epsilon/2g}$ is real and 
$\sigma=e^{i\beta}$, and so this excludes a direct connection with the GSG model. The deformation parameter 
of \cite{Delduc:2012qb,Delduc:2012mk,Delduc:2012vq,Delduc:2013fga,Delduc:2013qra} is 
\EQ{
\epsilon=\sin\beta\ .
}
which is restricted to $[0,1]$.
The deformation with real $q$ is also considered in \cite{Arutyunov:2013ega} ($g$ there is our $2g$ and $\nu=\epsilon$), where it is shown that the corresponding deformed action constructed in~\cite{Delduc:2013qra} is consistent with the S-matrix of \cite{Hoare:2011wr,Hoare:2012fc} (see also~\cite{Arutyunov:2012zt,Arutyunov:2012ai}) at leading order in perturbation theory. These deformations may also be related to those constructed in~\cite{Sfetsos:2013wia}.
 
If one na\"\i vely takes the deformed action of \cite{Delduc:2013qra} and takes $q$ to be a complex phase, then the action is no longer real. This is mirrored by the S-matrix in the vertex representation which is not unitary. However, unitarity at the level of the S-matrix can be restored by transforming from the vertex to the IRF representation \cite{Hoare:2013ysa}. In the relativistic limit $\sigma\to\infty$ this transformation is consistent with the topological quantization of soliton boundary conditions required to make sense of the WZ term in the Lagrangian formulation~\cite{Hollowood:2013oca}.
However, it remains to be seen how to implement the vertex-to-IRF transformation at the level of the action for generic values of $\sigma$.

\phantom{..}

\noindent
We would like to thank Ben Hoare for earlystage
collaboration on this project and for comments on the manuscript.
TJH is supported in part by the STFC grant ST/G000506/1.
JLM is supported in part by MINECO (FPA2011-22594), the Spanish Consolider-
Ingenio 2010 Programme CPAN (CSD2007-00042), Xunta de Galicia (GRC2013-024),
and FEDER.


\begin{thebibliography}{99}


{\small

\bibitem{Delduc:2012qb}
  F.~Delduc, M.~Magro and B.~Vicedo,
  JHEP {\bf 1208} (2012) 019
  [arXiv:1204.0766 [hep-th]].
  
\bibitem{Delduc:2012mk}
  F.~Delduc, M.~Magro and B.~Vicedo,
  Phys.\ Lett.\ B {\bf 713} (2012) 347
  [arXiv:1204.2531 [hep-th]].
  
\bibitem{Delduc:2012vq}
  F.~Delduc, M.~Magro and B.~Vicedo,
  JHEP {\bf 1210} (2012) 061
  [arXiv:1206.6050 [hep-th]].

\bibitem{Delduc:2013fga} 
  F.~Delduc, M.~Magro and B.~Vicedo,
  JHEP {\bf 1311}, 192 (2013)
  [arXiv:1308.3581 [hep-th]].
  
\bibitem{Delduc:2013qra}
  F.~Delduc, M.~Magro and B.~Vicedo,
  arXiv:1309.5850 [hep-th].

\bibitem{Mikhailov:2006uc} 
  A.~Mikhailov,
  Adv.\ Theor.\ Math.\ Phys.\  {\bf 14}, 1585 (2010)
  [hep-th/0609108].

\bibitem{Mikhailov:2007xr}
  A.~Mikhailov and S.~Schafer-Nameki,
  JHEP {\bf 0805} (2008) 075
  [arXiv:0711.0195 [hep-th]].
 
\bibitem{Schmidtt:2011nr} 
  D.~M.~Schmidtt,
  JHEP {\bf 1111}, 067 (2011)
  [arXiv:1106.4796 [hep-th]].

\bibitem{Magri:1977gn}
  F.~Magri,
  J.\ Math.\ Phys.\  {\bf 19} (1978) 1156.

\bibitem{Fokas:1997}
A.~S.~Fokas, P.~J.~Olver and P.~Rosenau,
Prog. in Nonlinear Differential Equations Appl. 26 (1997) 93-101.

\bibitem{Mikhailov:2005sy}
  A.~Mikhailov,
  J.\ Geom.\ Phys.\  {\bf 61} (2011) 85
  [hep-th/0511069].

\bibitem{Maltsev:2001}
A.~Ya.~Maltsev and S.~P.~Novikov, 
Physica D {\bf 156} (2001) 53-80.

\bibitem{DeSole:2013}
A.~De Sole and V.~G.~Kac, 
arXiv:1302.0148 [math-ph].

\bibitem{Miramontes:2008wt} 
  J.~L.~Miramontes,
  JHEP {\bf 0810}, 087 (2008)
  [arXiv:0808.3365 [hep-th]].

\bibitem{Bakas:1995bm}
  I.~Bakas, Q-H.~Park and H.~-J.~Shin,
  Phys.\ Lett.\ B {\bf 372} (1996) 45
  [hep-th/9512030].

\bibitem{Hollowood:2009tw} 
  T.~J.~Hollowood and J.~L.~Miramontes,
  JHEP {\bf 0904}, 060 (2009)
  [arXiv:0902.2405 [hep-th]].
  
\bibitem{Hollowood:2010dt} 
  T.~J.~Hollowood and J.~L.~Miramontes,
  JHEP {\bf 1104}, 119 (2011)
  [arXiv:1012.0716 [hep-th]].

\bibitem{Hollowood:2013oca} 
  T.~J.~Hollowood, J.~L.~Miramontes and D.~M.~Schmidtt,
  JHEP {\bf 1310}, 058 (2013)
  [arXiv:1306.6651 [hep-th]].

\bibitem{Hoare:2013ysa} 
  B.~Hoare, T.~J.~Hollowood and J.~L.~Miramontes,
  JHEP {\bf 1310}, 050 (2013)
  [arXiv:1303.1447 [hep-th]].
 
\bibitem{Dorey:2006dq} 
  N.~Dorey,
  J.\ Phys.\ A {\bf 39}, 13119 (2006)
  [hep-th/0604175].
   
\bibitem{Grigoriev:2007bu} 
  M.~Grigoriev and A.~A.~Tseytlin,
  Nucl.\ Phys.\ B {\bf 800}, 450 (2008)
  [arXiv:0711.0155 [hep-th]].
    
\bibitem{Hollowood:2011fq} 
  T.~J.~Hollowood and J.~L.~Miramontes,
  JHEP {\bf 1105}, 136 (2011)
  [arXiv:1104.2429 [hep-th]].

\bibitem{Goykhman:2011mq}
  M.~Goykhman and E.~Ivanov,
  JHEP {\bf 1109} (2011) 078
  [arXiv:1104.0706 [hep-th]].
 
\bibitem{Hoare:2011wr} 
  B.~Hoare, T.~J.~Hollowood and J.~L.~Miramontes,
  JHEP {\bf 1203}, 015 (2012)
  [arXiv:1112.4485 [hep-th]].

\bibitem{Hoare:2012fc} 
  B.~Hoare, T.~J.~Hollowood and J.~L.~Miramontes,
  JHEP {\bf 1210}, 076 (2012)
  [arXiv:1206.0010 [hep-th]].
  
\bibitem{Jackiw:1975im} 
  R.~Jackiw and G.~Woo,
  Phys.\ Rev.\ D {\bf 12}, 1643 (1975).

\bibitem{Arutyunov:2013ega} 
  G.~Arutyunov, R.~Borsato and S.~Frolov,
  arXiv:1312.3542 [hep-th].

\bibitem{Arutyunov:2012zt}
  G.~Arutyunov, M.~de Leeuw and S.~J.~van Tongeren,
  JHEP {\bf 1210} (2012) 090
  [arXiv:1208.3478 [hep-th]].

\bibitem{Arutyunov:2012ai}
  G.~Arutyunov, M.~de Leeuw and S.~J.~van Tongeren,
  JHEP {\bf 1302} (2013) 012
  [arXiv:1210.8185 [hep-th]].

\bibitem{Sfetsos:2013wia}
  K.~Sfetsos,
  Nucl.\ Phys.\ B {\bf 880} (2014) 225
  [arXiv:1312.4560 [hep-th]].

}

\end{thebibliography}
\end{document}